\documentclass[onecolumn,showpacs,aps,prd,superscriptaddress, amsmath,amssymb]{revtex4}
\usepackage{graphicx}
\usepackage{dcolumn}
\usepackage{bm}
\usepackage{amsmath, amssymb, graphics, setspace}
\usepackage{latexsym,epsf,rotate}
\usepackage{hyperref}
\usepackage{xcolor}
\usepackage{epsfig, epstopdf}

\def\beq{\begin{equation}} 
\def\eeq{\end{equation}}
\def\bea{\begin{eqnarray}}
\def\eea{\end{eqnarray}}
\def\eqref#1{eq.~(\ref{eq:#1})}

\def\nn{\nonumber}   

\begin{document}
\hspace{12.5cm} IFJPAN-IV-2024-8 
\vspace{0.5cm}
\title{$\tau$-lepton pair spin in proton-proton LHC collisions for anomalous dipole moments }
\author{A.Yu.~Korchin} \email{korchin@kipt.kharkov.ua}
\affiliation{NSC Kharkiv Institute of Physics and Technology, 61108 Kharkiv, Ukraine}
\affiliation{V.N.~Karazin Kharkiv National University, 61022 Kharkiv, Ukraine}
\affiliation{Institute of Physics, Jagiellonian University, ul. Lojasiewicza 11, 30-348 Krakow, Poland}
\author{E.~Richter-Was} \email{Elzbieta.Richter-Was@uj.edu.pl}
\affiliation{Institute of Physics, Jagiellonian University, ul. Lojasiewicza 11, 30-348 Krakow, Poland}
\author{Yu.~Volkotrub  }  \email{yuriy.volkotrub@uj.edu.pl}
\affiliation{Institute of Physics, Jagiellonian University, ul. Lojasiewicza 11, 30-348 Krakow, Poland}
\author{Z.~Was} \email{z.was@ifj.edu.pl}
\affiliation{Institute of Nuclear Physics Polish Academy of Sciences, PL-31342 Krakow, Poland}


\begin{abstract}
The spin correlations in the pair of $\tau$ leptons produced in the high-energy proton-proton collisions
are studied. The invariant mass of the $\tau$-lepton pair is chosen close to $Z$-boson mass, in these conditions $Z$-boson exchange gives the dominant contribution to the Drell-Yan mechanism.  
The interaction of $Z$ boson with $\tau$ lepton, in addition to the Standard Model couplings, can include 
weak anomalous magnetic and electric dipole moments (or form-factors) of the $\tau$ lepton. Dependence of elements of the $pp$ spin-correlation matrix on the weak anomalous moments is determined for various regions of the rapidity of the $\tau$-lepton pair.  Based on this calculation, we construct a semi-realistic observable built on the momenta of pions in the $\tau^\mp \to \pi^\mp \nu_\tau$ and 
$\tau^\pm \to \rho^\pm \nu_\tau \to \pi^\pm \pi^0 \nu_\tau$ decay channels, and show that this observable 
can be sensitive to the weak anomalous dipole moments.
\end{abstract}


\maketitle

\setcounter{footnote}{0} 

\section{Introduction}
\label{sec:Introduction}

One of the interesting topics of present-day phenomenology relates to measurement 
or establishing limits on the muon anomalous dipole moments, magnetic and electric ones. Numerous papers were devoted to that subject (see~\cite{Jegerlehner:2017gek, Aoyama:2020ynm} and references therein), and experimental 
measurements~\cite{Muong-2:2021ojo,  Muong-2:2021vma, Muong-2:2023cdq} 
may indicate that deviation of the measured value of anomalous magnetic moment from the Standard Model (SM) prediction is at the edge of the discovery. Independently, if this is the case, one should prepare alternative or supplementary measurements and necessary for their interpretation calculations and software tools.

The production and decay of $\tau$-lepton pair offer interesting research direction, because anomalous magnetic dipole moments due to possibly enhanced contributions from New Physics (NP) models 
are expected to be proportional to the mass squared of the lepton, and are therefore enhanced for the $\tau$ lepton by a factor of 280 compared to the muon. This makes these studies important and of contemporary interest.

The LHC experiments \cite{ATLAS:2022ryk, CMS:2022arf, CMS:2024skm}, taking data now, offer immediate opportunity for the measurements. The $\tau$ lepton decays and the $\tau$-decay products are observed and provide access to the physics information embedded in the $\tau$-lepton pair spin state.
Neither incoming partons in the hard scattering process nor all $\tau$-decay products (due to neutrinos escaping detection) are available for direct measurement. To simplify the task of observable build and
of measurement interpretation, tools like {\tt TauSpinner} 
\cite{Richter-Was:2018lld,Przedzinski:2018ett,Richter-Was:2020jlt}
can be helpful. Since Ref.~\cite{Banerjee:2023qjc}, the tool can be used to calculate event weights introducing anomalous dipole moments.

However, these papers do not address the actual impact of dipole moments
on $\tau$-decay products distributions. In the present paper, we address the topic in the context of the $pp$ collisions in the conditions of the LHC, which was not covered previously.
The consideration is focused on the region of $\tau$-pair invariant mass around $Z$-boson 
mass. In these conditions, the $Z$-boson exchange mechanism of the Drell-Yan production 
of the $\tau$ pair is the dominant one, and the coupling of $Z$ boson to $\tau$ leptons may carry 
information on weak anomalous magnetic and/or electric dipole moments of the $\tau$ lepton.   
 
In Section \ref{sec:kinematics} we describe differential cross section in terms of 
the spin-correlation matrix in the high-energy proton-proton collisions. 
In Section \ref{sec:results_for_rij} 
some results for elements of effective proton-proton spin-correlation matrix are presented in 
different kinematical regimes.  
Section \ref{sec:decays} is devoted to the definition of a prototype observable that is sensitive to weak anomalous dipole moments. It is distinct from the one used in {\tt TauSpinner} 
algorithm because it relies on detectable momenta only; it does not use the momenta of $\tau$ leptons. 
Conclusions are given in Section \ref{sec:summary}. 


\section{Differential cross section of $p + p \to Z + \ldots \to  \tau^- + \tau^+ + \ldots$ }
\label{sec:kinematics}

In this section, we present a differential cross section of the reaction 
$p + p \to Z + \ldots \to  \tau^- + \tau^+ + \ldots$  
as a function of the $\tau$-lepton rapidities $y_-, \, y_+$ and $\tau$-pair invariant mass (squared) 
$M^2= (p_- +p_+)^2 = x_1 x_2 s$, where $s$ is invariant energy 
squared in the $pp$ collision.  Here $x_1$ and $x_2$ are fractions of the proton momenta $p_1$ and $p_2$ 
carried by partons $f$ and $\bar{f}$, respectively.   

The differential cross section of the process $p + p \to \tau^- + \tau^+ + \ldots$ is written as 
(see, e.g. \cite{Peskin:1995ev}, Ch.~17.4)
\begin{equation}
\frac{d \sigma}{d x_1 \, dx_2 \,  d t} (p \, p \to \tau^-  \tau^+ + \ldots) = 
\sum_{f=q, \bar{q}}  \, F_f (x_1) F_{\bar{f}}(x_2)  \frac{d \sigma }{d t} (f  \bar{f} \to \tau^-  \tau^+),  
\label{eq:CS_1}
\end{equation}
where $F_{f, \, \bar f}(x)$ are the parton distribution functions (PDF)s, which also depend on an energy scale 
$Q$ (not indicated explicitly).

The  total four-momentum of the $\tau$ pair  in the parton-parton 
process $f(x_1 p_1) + \bar{f}(x_2p_2) \to \tau^- (p_-) + \tau^+ (p_+)$ is 
\begin{eqnarray}
\label{eq:total_q} 
&& q^\mu= (q^0, \, \vec{q}), \qquad \vec{q}=(0, \, 0, \, q), \nn  \\
&& q^0 = E_- + E_+ = \frac{\sqrt{s}}{2}   \, (x_1 +x_2),  \nn \\
&& q = p_{- \parallel} + p_{+ \parallel} =  \frac{\sqrt{s}}{2} \, (x_1 - x_2).  
\end{eqnarray}
Hereafter the mass of the proton is neglected.  The fractions of the proton momenta
\begin{eqnarray}
x_1 = \frac{M}{\sqrt{s}} \, \exp{(Y_Z)}, \qquad x_2 =  \frac{M}{\sqrt{s}} \, \exp{(-Y_Z)} 
\label{eq:x1_x2}
\end{eqnarray}
are expressed through the invariant mass and rapidity of the $\tau$ pair 
\begin{equation}
Y_Z =  \ln \left( \frac{q^0 + q}{M} \right) =  \frac{1}{2} \ln \left( \frac{x_1}{x_2} \right). 
\label{eq:rapidity_Y}
\end{equation}

To include angular dependence of the hard process $f \, \bar{f} \to \tau^- \tau^+$ 
we introduce the longitudinal rapidities of $\tau^-$ and $\tau^+$ via the relations 
\cite{Peskin:1995ev}:
\begin{eqnarray}
&& E_- = p_\perp \cosh (y_-),    \qquad   E_+ = p_\perp \cosh (y_+), \nn \\        
&&  p_{- \parallel} = p_\perp \sinh (y_-),  \qquad  p_{+ \parallel} = p_\perp \sinh (y_+),  \nn \\
&& \vec{p}_{- \perp} = \vec{n}_\perp \, p_\perp, \qquad  \qquad \;
\vec{p}_{+ \perp} = - \vec{n}_\perp \, p_\perp, 
\label{eq:rapidities_y-y+}
\end{eqnarray}
where ``parallel'' (``perpendicular'') means longitudinal (transverse) to the beam components of the $\tau^\mp$ momenta, $\vec{n}_\perp$ is the unit vector in the transverse direction, 
and we assume that initial partons do not have transverse momenta. 

It is straightforward to relate rapidities of $\tau^-$ and $\tau^+$ with the lepton pair rapidity $Y_Z$ and 
variable $y^*$, which has the meaning of  the $\tau^-$ rapidity in the center-of-mass (c.m.) frame of the 
$\tau^- \tau^+$ pair. Namely, 
\begin{equation}
y_- = Y_Z + y^*, \qquad y_+ = Y_Z -y^*.
\label{eq:Y_y*}
\end{equation}
The rapidity of $\tau^+$ in the c.m. frame is $\, - y^*$~\footnote{Kinematic variables in the $\tau^- \, \tau^+$ c.m. frame are indicated with superscript $*$. }.
Note that the c.m. angle $\theta^*$ between the parton $f$ and $\tau^-$ is determined from  
\begin{equation}
\sin \theta^* = \frac{1}{\cosh(y^*)}, \qquad \cos \theta^* = \tanh(y^*) .
\label{eq:CM_angle}
\end{equation}
The transverse momentum of $\tau$'s is $p_\perp =  \frac{1}{2} M \sin \theta^*$ 
and the invariant momentum transfer $t$ of Eq.~({\ref{eq:CS_1}) is equal to  
$ t= -\frac{1}{2} M^2 (1 - \cos \theta^*)$.

We will choose the variables $Y_Z$, $y^*$, and $M^2$ as independent ones. 
Then the triply differential cross section can be written as  
\begin{equation}
\label{eq:CS_2}  
\frac{d \sigma}{dY_Z dy^* d M^2 } (p \, p \to \tau^-  \tau^+ + \ldots) =
\frac{x_1 x_2}{ 2 \cosh^2(y^*)}  \sum_{f = q, \bar{q}}  F_f (x_1) F_{\bar{f}}(x_2) 
\frac{d \sigma}{d t} (f  \bar{f} \to \tau^-  \tau^+),  
\end{equation}
where the factor  $x_1 x_2$ originates from the Jacobian due to change of variables from 
$(x_1,  x_2,  t)$ to $(Y_Z,   y^*,  M^2)$.  The cross section of the quark-antiquark production of the polarized $\tau$ leptons is~\cite{Banerjee:2023qjc} 
\begin{equation}
\frac{d \sigma}{d t} (q \bar{q} \to \tau^- \tau^+) =  \frac{\beta}{16 \pi M^4} \, \sum_{i, j =1}^4 R^{(q \bar{q})}_{i, j}(M, \theta^*) 
\, S^{-}_i S^{+}_j. 
\label{eq:CS_parton}
\end{equation}
Here $S^\mp_{i} \equiv (\vec{S}^{\, \mp}, \, 1)$,  $\vec{S}^\mp$ are the 
polarization vectors of $\tau^\mp$ in their corresponding rest frames. 
The Cartesian components of vectors $\vec{S}^{\, \mp}$ are defined in a system  
with the axes $(\hat{x}^\prime,  \hat{y}^\prime, \hat{z}^\prime)$, where $\hat{z}^\prime$ is 
along the c.m. momentum of $\tau^-$, plane $\hat{x}^\prime \hat{z}^\prime$ is spanned 
on the c.m. momenta of $\tau^-$ and $q$, and $\hat{y}^\prime=\hat{z}^\prime \times \hat{x}^\prime$.  
The normalized spin-correlation matrix is expressed in terms of the coefficients 
$R^{(q \bar{q})}_{i, j} (M, \theta^*) $ as  
$ r_{i, j}^{(q \bar{q})} \equiv R_{i, j}^{(q \bar{q})}/R_{44}^{(q \bar{q})}$. 
Besides,  in Eq.~(\ref{eq:CS_parton}) $\beta = (1 - 4 m_\tau^2/M^2)^{1/2} \approx 1$ 
is the $\tau$-lepton velocity in the c.m. frame.    


\subsection{Analysis of kinematics}
\label{subsec:kinematics}

A typical configuration of $\tau$-lepton momenta in laboratory frame is shown in Fig.~\ref{fig:kinematics}.
\begin{figure}[!htb]
\begin{center}
\includegraphics[height=90pt]{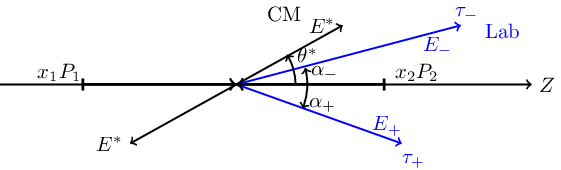} 
\end{center}
\caption{Configuration of $\tau^+$ and $\tau^-$ momenta in the laboratory and c.m. frames.}
\label{fig:kinematics}
\end{figure} 
Here $\alpha_+$ and $\alpha_-$ are the angles of $\tau^+$ and $\tau^-$ with respect to the
proton beam axis. They are related to the corresponding rapidities  
\begin{eqnarray}
&& \cos \alpha_\pm = \frac{p_{\pm \parallel}}{E_\pm} = \tanh(y_\pm),  \nn \\  
&& y_\pm = \tanh^{-1} (\cos \alpha_\pm ) = 
\frac{1}{2} \ln \Bigl( \frac{1 + \cos \alpha_\pm }{1 - \cos \alpha_\pm } \Bigr) .
\label{eq:alpha}
\end{eqnarray}
Note that   $0 \le \alpha_\pm \le \pi/2$ if $y_\pm \ge 0$, and $\pi/2 \le \alpha_\pm \le \pi$ if 
$y_\pm \le 0$. 

Let us analyze the scattering angle $\theta^*$ which determines
the $q \bar{q} \to \tau^- \tau^+$ spin-correlation matrix in the c.m. frame.  
  
(i) If $x_1$ corresponds to quark and $x_2$ to antiquark, then $\theta^*$ is the angle between 
the quark and $\tau^-$ lepton.  
In the interval $0 \le \theta^* \le \pi$ the sign of $\sin \theta^*$ is positive, 
while the sign of $\cos \theta^*$ depends on the sign of $y^*$.
 
(ii) If $x_1$ corresponds to antiquark and $x_2$  to quark, then one needs to change the angle 
$\theta^* \to \pi - \theta^*$ in the matrix $R_{i, j}^{(q \bar{q})}(M, \theta^*)$, and  perform additional rotation on the latter (see further). The corresponding cross section is denoted below with ``tilde''.

The cross section taking into account both possibilities is written in the form
\begin{eqnarray}
\label{eq:CS_3}
 \frac{d \sigma}{dY_Z dy^* d M^2 }  (p \, p \to  \tau^-  \tau^+ + \ldots) &=& 
 \frac{x_1 x_2}{ 2 \cosh^2(y^*)}     
\sum_{q = u, d, s, \ldots } \Bigl[
 F_q (x_1) \, F_{\bar{q}}(x_2)  \, \frac{d \sigma }{d t} (q \bar{q} \to \tau^- \tau^+)  \nn   \\
&& +  F_{\bar{q}} (x_1) \,  F_{q}(x_2)  \, \frac{d \tilde{\sigma}}{d t} (q \bar{q} \to \tau^- \tau^+) 
	 \Bigr].
\end{eqnarray}


\subsection{Kinematics for  $\tau$ spin quantization frames}
\label{subsec:quantization frames}

For the spin correlation matrix to be useful, we need to define all axes of 
the corresponding reference frames for $\tau^-$ and $\tau^+$ separately.

Then, we can re-write and extend Eq.~(\ref{eq:CS_3}) to account for the decays 
$\tau^- \to X^- \nu_\tau$ and $\tau^+ \to X^{+} \bar{\nu}_\tau$:
\begin{eqnarray}
\frac{d \sigma  }{dY_Z dy^* d M^2 d\Omega_- d\Omega_+}  (p \, p   \to  X^- \,X^+ \, 
 \nu_\tau \, \bar{\nu}_\tau + \ldots)  & = &  
 \frac{\beta  x_1 x_2}{32 \pi M^4 \cosh^2(y^*) } \nn  \\
 && \times \sum_{q = u, d, s, \ldots }  \Bigl[
 F_q (x_1) \,  F_{\bar{q}}(x_2)  \sum_{i, j =1}^4 R_{i, j}^{(q \bar q)} (M,  \theta^*) \,  h_i^{-}  h_j^+   \nn \\ 
&& + F_{\bar{q}} (x_1) \,  F_{q}(x_2)  \sum_{i, j =1}^4   \tilde{R}_{i, j}^{(q  \bar{q})} (M,  \pi -\theta^*) 
\,   h_i^{-}  h_j^+  \Bigr],
\label{eq:main}
\end{eqnarray}
where $\Omega_-, \,  \Omega_+$ are the solid angles of the final hadrons $X^-, \, X^{+}$.  
In  Eq.~(\ref{eq:main}) $h^-_i$ and $h^+_j$ are the spin polarimetric vectors of 
$\tau^-$ and $\tau^+$ decays, respectively, which depend on the $\tau$-decay products momenta. 

The first sum in the brackets of Eq.~(\ref{eq:main}) corresponds to contribution where the 
laboratory $\hat z$ axis corresponds to the direction of the incoming quark $q$. 
For the second part, the laboratory $\hat{z}$ axis is along the direction of antiquark $\bar{q}$. 
Because we are interested in the spin correlations, and not in the unpolarized cross section, the second hard-scattering frame needs to be properly oriented.  This results in 
\begin{equation}
\tilde{R}_{i, j}^{(q \bar{q})}(M, \pi - \theta^*)  \equiv   
\hat{O}_{z^\prime} (\pi) \, R_{i, j}^{(q \bar{q})}(M, \pi- \theta^*)  
=  \varepsilon_i \, \varepsilon_j \,  R_{i, j}^{(q \bar{q})}(M, \pi- \theta^*), 
\label{eq:def_tilde_R}
\end{equation}
where $ \hat{O}_{z^\prime} (\pi) $ is rotation around $\hat{z}^\prime$ axis by the angle $\pi$. Further,  
$\varepsilon_i = -1$ for $i=1,2$ and $\varepsilon_i = +1$ for $i=3, 4$.

To calculate event weight we sum the spin-correlation coefficients in the $q \, \bar{q} \to \tau^- \tau^+$ process over the parton flavors  
\begin{equation}
\label{eq:sum_flavors}
wt_{prod} \, wt_{spin} = \sum_{i, j =1}^4 \,  R_{i,j}^{(pp)} \, h_i^{-} \,  h_j^+  
=  R_{4 4}^{(pp)} \Bigl(1 +  \sum_{i=1}^3 \, r_{i, 4}^{(pp)} h_i^- + 
\sum_{j=1}^3 \, r_{4, j}^{(pp)} h_j^+ 
+  \sum_{i, j =1}^3 \,  {r}_{i,j}^{(pp)} \, h_i^{-} \, h_j^+   \Bigr),  
\end{equation}
where 
\begin{equation}
\label{eq:Rij}
 R_{i,j}^{(pp)}  \equiv  x_1 x_2  \sum_{q=u,  d,  s, \ldots}  \Bigl[
    F_q (x_1) \,  F_{\bar{q}}(x_2) R_{i, j}^{(q \bar q)} (M,  \theta^*)     
+  F_{\bar{q}} (x_1) \,  F_{q}(x_2)   \tilde{R}_{i, j}^{(q \bar{q})} (M,  \pi -\theta^*) \Bigr] 
\end{equation}
for $i,j=1,2,3,4$, and define elements of effective proton-proton spin-correlation matrix $r_{i,j}^{(pp)}$  
and $\tau^-$ ($\tau^+$) polarization $r_{i,4}^{(pp)}$ ($r_{4,j}^{(pp)}$) in Eq.~(\ref{eq:sum_flavors}) as
\begin{equation}
r_{i,j}^{(pp)} =  \frac{R_{i,j}^{(pp)}}{R_{4 4}^{(pp)}}, \qquad  
r_{i,4}^{(pp)} =  \frac{R_{i, 4}^{(pp)}}{R_{4 4}^{(pp)}}, \qquad 
r_{4,j}^{(pp)} =  \frac{R_{4, j}^{(pp)}}{R_{4 4}^{(pp)}} 
\label{eq:rij_definition}
\end{equation}
with $ i, j = 1,2,3$. These elements depend on the variables  $Y_Z, \, y^*$ and $M^2$. We further omit 
superscript (pp) for brevity. 

The definition of the anomalous magnetic $X(M^2)$ and electric $Y(M^2)$ dipole form-factors, and their relation to the corresponding weak dipole moments at $M=M_Z$ is given in~\cite{Banerjee:2023qjc}.   
Note that in the calculation only the linear in $X(M^2)$ and $Y(M^2)$ contributions are included, as one can expect their values to be very small. Indeed, the experiment ALEPH~\cite{ALEPH:2002kbp} determined the following constraints
\begin{eqnarray}
&& |{\rm Re}(\mu_\tau) |_{exp } < 1.14  \times 10^{-3},   \qquad
 |{\rm Im}(\mu_\tau) |_{exp} < 2.65 \times 10^{-3}, \nn \\
&& |{\rm Re}(d_\tau) |_{exp} <  0.91  \times 10^{-3}, \qquad
|{\rm Im}(d_\tau) |_{exp} <   2.01 \times 10^{-3},
\label{eq:ALEPH}
\end{eqnarray}  
while the SM prediction \cite{Bernabeu:1994wh} for the weak anomalous magnetic moment 
is~\footnote{Our dipole moments are related to $\mu_\tau$ and $d_\tau$ via the relations:
$X(M_Z^2) =  \mu_\tau  \sin 2 \theta_W,  \;  Y(M_Z^2) = d_\tau  \sin 2 \theta_W$, where  
 $\theta_W$ is the weak mixing angle and $\sin 2 \theta_W \approx 0.843$.} 
\begin{equation} 
\mu_{\tau, \, SM} = - (2.10 + i \, 0.61)  \times 10^{-6}. 
\label{eq:mu_SM}
\end{equation}  

The detailed definitions of axes of the $\tau$ reference frames are not necessary 
for the  formula (\ref{eq:sum_flavors}) phenomenology to be discussed.
We will return to this point later, in fact, we will rely on the frames built from secondary decays momenta.


\section{Numerical results on spin density matrix} 
\label{sec:results_for_rij}

In this section, we discuss numerical results on the spin-correlation matrix averaged over  
PDFs in Eqs.~(\ref{eq:Rij}). The PDFs ``MSTW 2008'' from Ref.~\cite{Martin:2009iq} are used in the calculation.      

Let us start a discussion of the numerical results for the elements of the matrix in 
Eqs.~(\ref{eq:rij_definition}).
There are two interesting kinematic regimes. The first corresponds to the small rapidity of the $\tau$-lepton pair, and the second corresponds to the situation, where the $\tau$ pair is strongly boosted. 
In the later case the quark is more probable to come from the first of the  colliding protons.
For example, from Table~\ref{tab:kinematics_1} and also from behavior of PDFs
given in Fig.~\ref{fig:PDFs} it is clear, that at large rapidity $Y_Z = 4$ the parton carrying fraction $x_1$ of the proton momentum is usually the valence quark, while the other parton is the sea antiquark.

In the figures below, we choose the proton-proton energy $\sqrt{s}=13.6$ TeV, the $\tau^- \tau^+$ pair 
invariant mass $M=M_Z$, and fix the total $\tau$-pair rapidity $Y_Z$. To completely determine kinematics one needs to fix also the c.m. rapidity $y^*$.   
The latter is chosen to correspond to the c.m. angle $\theta^* = 60^\circ$. 
Values of kinematic variables for the three values $Y_Z =0, \, 2, \, 4$ are shown 
in Table~\ref{tab:kinematics_1}.
\begin{table}[!tbh]  
\begin{center}
\caption{Rapidity $y^*$, rapidities of $\tau^-$ and  $\tau^+$ leptons $y_-$ and $y_+$,  
parton fractions $x_1, \, x_2$,  angles $\alpha_-$ and $\alpha_+$  of $\tau^-$ and $\tau^+$ in laboratory frame, 
c.m. angle $\theta^*$, transverse momentum $p_\perp$ of $\tau$ leptons 
and their energies $E_-, \, E_+$ in laboratory frame. } 
\vspace{0.1cm}
\begin{tabular}{l | l | l | l    }       
\hline \hline
 variable                    & \ $Y_Z=0$                   & \ $Y_Z=2$                     & \ $Y_Z=4$  \\
\hline
$y^*$                 & \ 0.54931  & \ 0.54931 & \ 0.54931           \\
$y_-$                   & \ 0.54931              & \ 2.54931     & \  4.54931             \\
$y_+$                   & \ -0.54931              & \  1.45069    &  \  3.45069             \\ 
$x_1$                 & \ 0.00671         &  \ 0.04954       & \ 0.36608              \\
$x_2$                  & \ 0.00671         & \  0.00091   &  \  0.00012             \\
$\alpha_-$ (deg)  &  \  60              &  \ 8.93556   &  \ 1.21171             \\
 $\alpha_+$ (deg)  & \   120              & \  26.3848   &  \  3.63404            \\
$\theta^*$ (deg)  & \ 60              & \ 60           & \ 60             \\ 
$ p_\perp$ (GeV)          & \   39.4854           &  \ 39.4854        &  \  39.4854  \\
$E_-$  (GeV)        &    \ 45.5938             & \  254.214               & \  1867.21        \\
 $E_+$  (GeV)        &  \  45.5938            & \  88.8516              &  \  622.961       \\
\hline \hline
\end{tabular}
\label{tab:kinematics_1}
\end{center}
\end{table}



\begin{figure}[!htb]
\begin{center}
  \includegraphics[height=135pt]{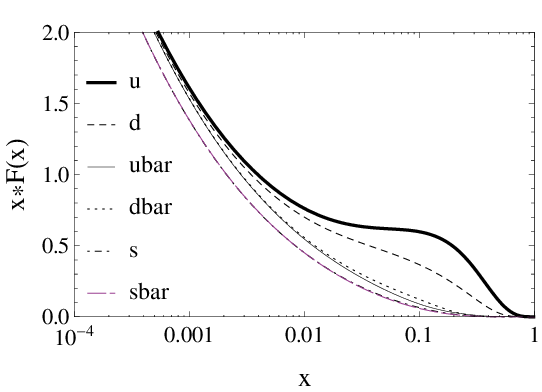}
\end{center}
\caption{Parton distribution functions MSTW 2008 at the energy scale $Q=M_Z$.}
\label{fig:PDFs}
\end{figure}

\begin{figure}[!htb]
\begin{center}
  \includegraphics[height=100pt]{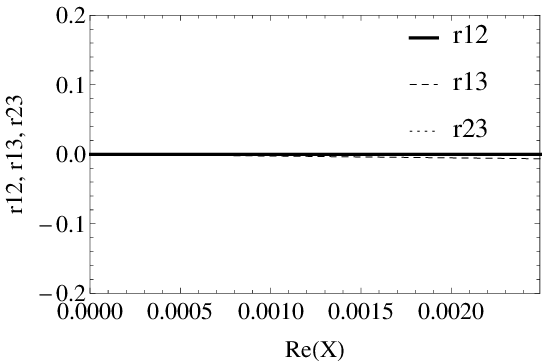} 
  \includegraphics[height=100pt]{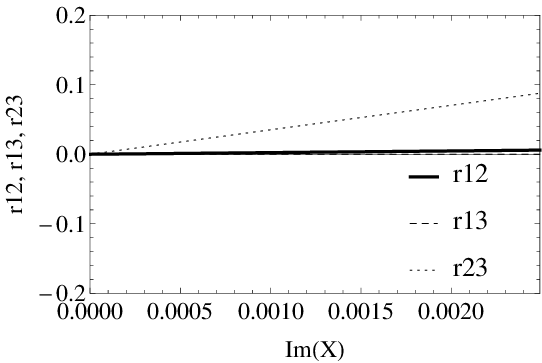} 	\\
  \includegraphics[height=100pt]{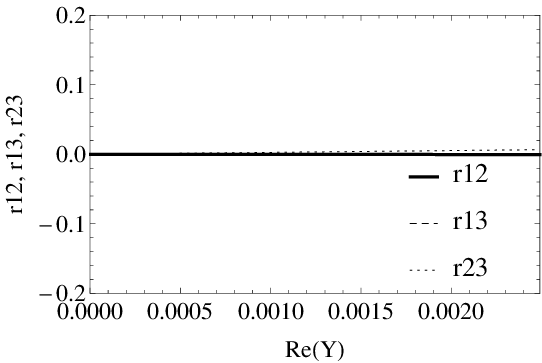} 
		\includegraphics[height=100pt]{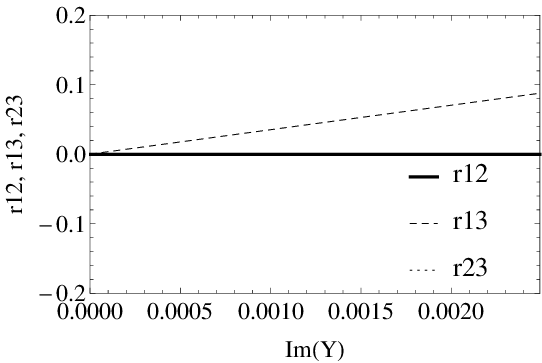} \\
  \includegraphics[height=100pt]{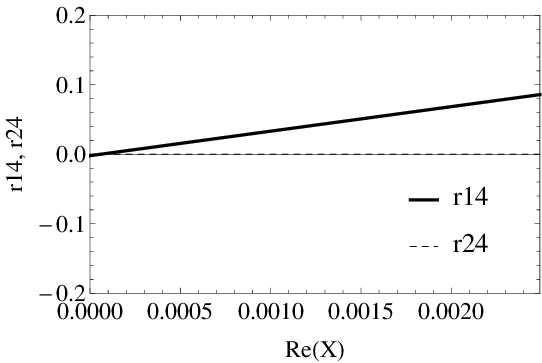} 
  \includegraphics[height=100pt]{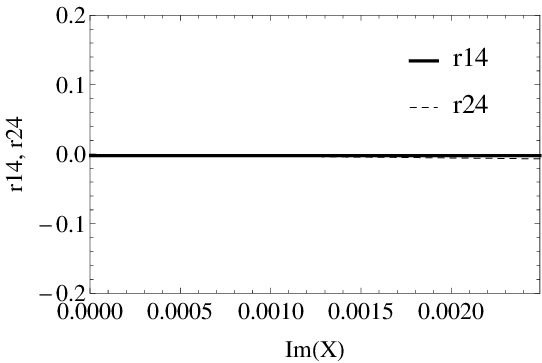} \\
  \includegraphics[height=100pt]{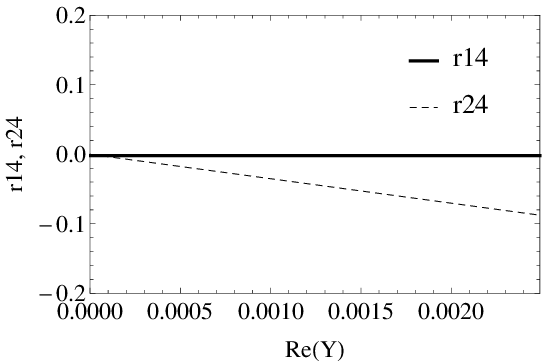} 
	\includegraphics[height=100pt]{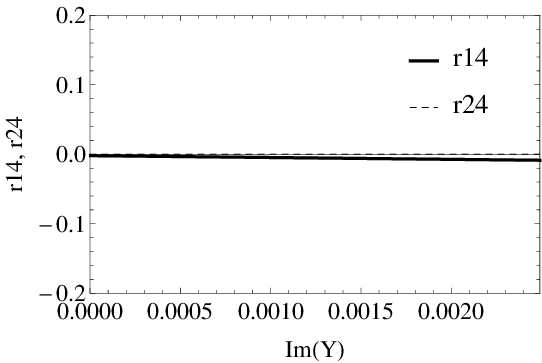}  
\end{center}
\caption{Dependence of $r_{i,j}$ on the real and imaginary parts of the weak magnetic and electric dipole moments. 
Proton-proton energy is $13.6$ TeV, $\tau$-pair invariant mass is equal to $M_Z$,  
rapidity $Y_Z=0$ and c.m. angle is $60^\circ$. The contribution to the cross 
section of the quark with the momentum fraction $x_1$ (and correspondingly antiquark with fraction $x_2$) is $55.4 \%$, and of the antiquark with fraction $x_1$ (and correspondingly quark with fraction $x_2$) -- $44.6 \%$. Elements $r_{i,j}$ which are not displayed in the figure are consistent with zero.}
 \label{fig:Y_Z=0}
\end{figure} 

\begin{figure}[!htb]
\begin{center}
\includegraphics[height=100pt]{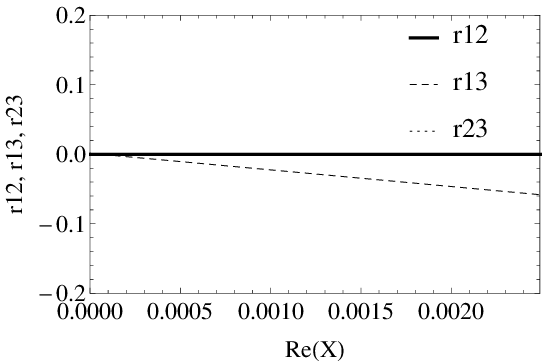}
\includegraphics[height=100pt]{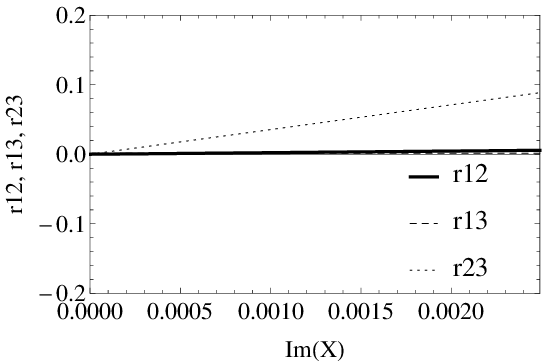} \\	
\includegraphics[height=100pt]{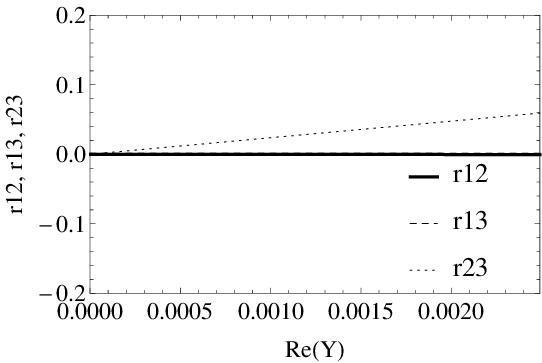}
 \includegraphics[height=100pt]{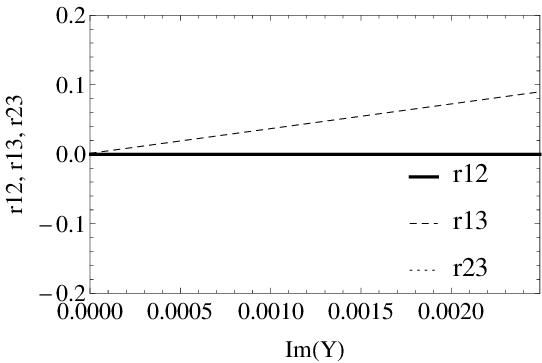} \\
  \includegraphics[height=100pt]{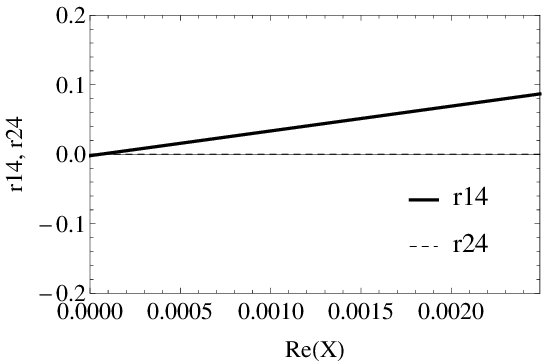}
 \includegraphics[height=100pt]{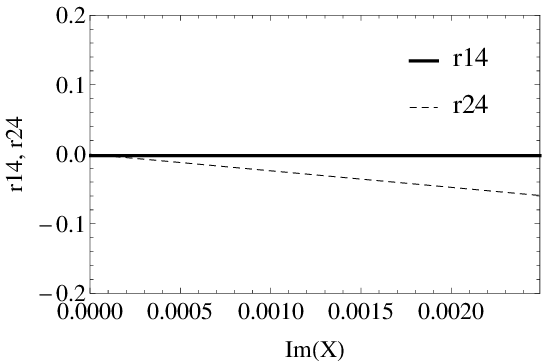} \\	
 \includegraphics[height=100pt]{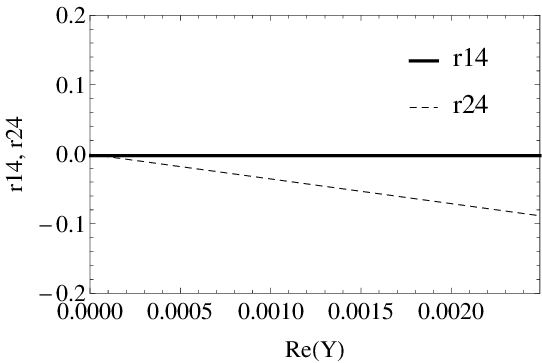}
 \includegraphics[height=100pt]{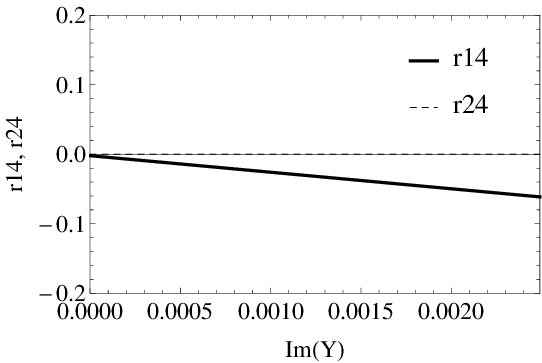} 
\end{center}
\caption{The same as in Fig.~\ref{fig:Y_Z=0} but for rapidity $Y_Z=2$. The contribution to the cross 
section of the quark with the momentum fraction $x_1$ (and correspondingly antiquark with fraction $x_2$) 
is $72.5 \%$, and of the antiquark with fraction $x_1$ (and correspondingly quark with fraction $x_2$) 
-- $27.5 \%$.}
 \label{fig:Y_Z=2}
\end{figure} 

\begin{figure}[!htb]
\begin{center}
  \includegraphics[height=100pt]{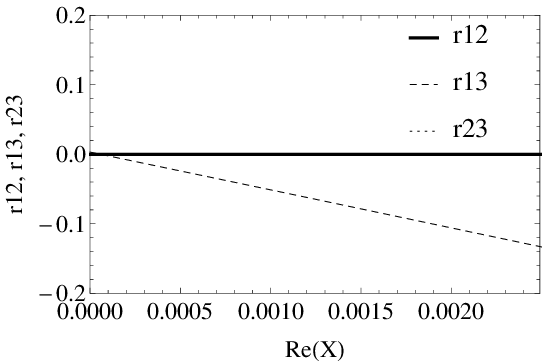}
  \includegraphics[height=100pt]{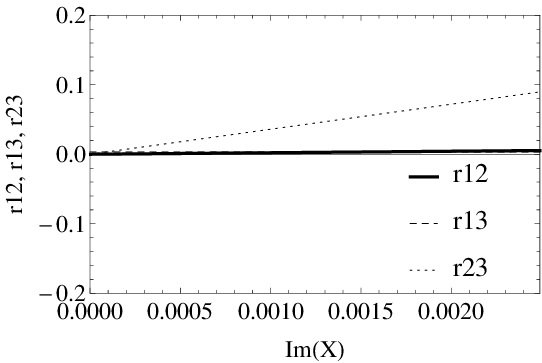} \\	
  \includegraphics[height=100pt]{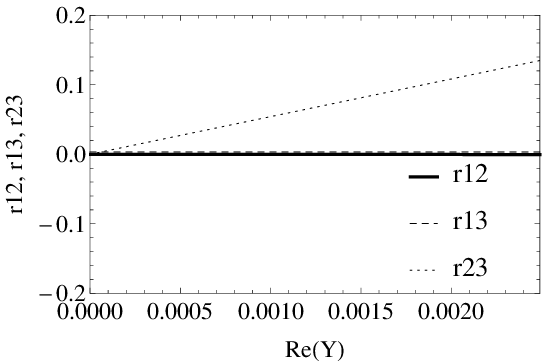}
  \includegraphics[height=100pt]{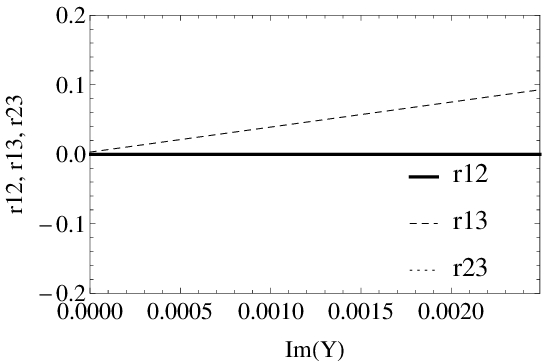} \\
  \includegraphics[height=100pt]{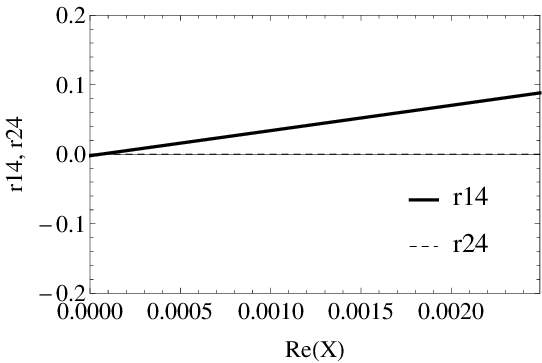}
  \includegraphics[height=100pt]{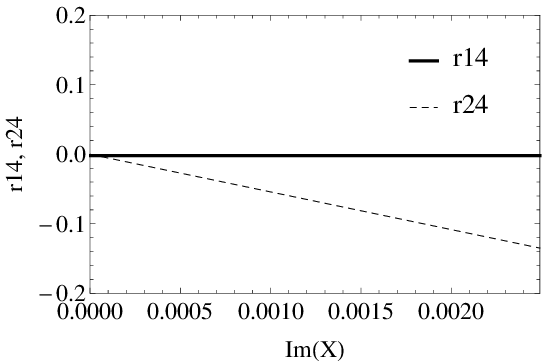} \\	
  \includegraphics[height=100pt]{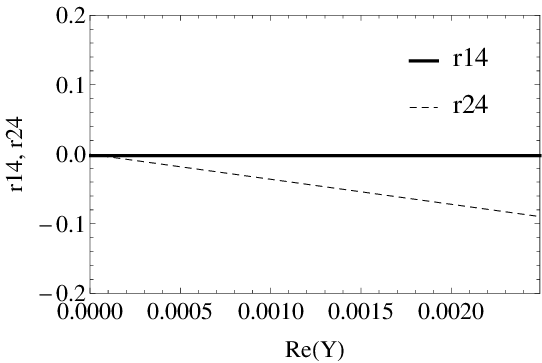}
  \includegraphics[height=100pt]{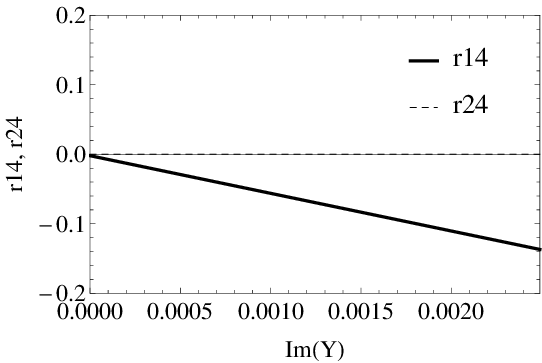} 
\end{center}
\caption{The same as in Fig.~\ref{fig:Y_Z=0} but for rapidity $Y_Z=4$. The contribution to the cross section of the quark with the momentum fraction $x_1$ (and correspondingly antiquark with fraction $x_2$) is 
$97.8 \%$, and of the antiquark with fraction $x_1$ (and correspondingly quark with fraction $x_2$) 
-- $2.2 \%$. }
 \label{fig:Y_Z=4}
\end{figure} 


In Figs.~\ref{fig:Y_Z=0}, \ref{fig:Y_Z=2} and \ref{fig:Y_Z=4} we  present  
dependence of the non-diagonal elements $r_{12}$, $r_{13}$, $r_{2 3}$, $r_{1 4}$ and $r_{2 4}$
on values of the real and imaginary parts of the weak anomalous magnetic and electric dipole moments 
$X$ and $Y$, respectively. The diagonal elements of the spin-correlation matrix, 
$r_{11}$, $r_{22}$ and $r_{33}$, are less sensitive to $X$ and $Y$ and are not shown.   

The element $r_{34}$ is equal to the longitudinal polarization of the $\tau^-$ lepton, and its value depends 
on the rapidity $Y_Z$ due to the different composition of the parton flavors (an average in Eq.~(\ref{eq:Rij})). In particular, in the SM   
\begin{eqnarray}
&& r_{34} = -0.150 \quad {\rm for } \quad Y_Z=0, \nn \\
&& r_{34} = -0.386 \quad {\rm for } \quad Y_Z=2, \nn \\
&& r_{34} = -0.725 \quad {\rm for } \quad Y_Z=4, 
  \label{eq:r34}
\end{eqnarray}  
and $r_{34}$ depends weakly on the anomalous dipole moments. For this reason, it is  
not shown in Figs.~ \ref{fig:Y_Z=0}--\ref{fig:Y_Z=4}.  

One can see that some elements of $r_{i, j}$ and $r_{i, 4}$ are sensitive to dipole moments. 
In particular, it is seen from Fig.~\ref{fig:Y_Z=0} that $r_{14}$ and $r_{24}$ are sensitive to ${\rm Re}(X)$ 
and ${\rm Re}(Y)$, while $r_{23}$ and $r_{13}$ are
sensitive to ${\rm Im}(X)$ and ${\rm Im}(Y)$. This may be convenient for separating magnetic and electric moment contributions, as well as their real and imaginary parts.

The pattern at larger rapidities in Figs.~\ref{fig:Y_Z=2} and \ref{fig:Y_Z=4} becomes more complex, although sensitivity of $r_{i, j}$ to dipole moments is enhanced. 

One can see that regimes of small and large (but still available for measurements at about $Y_Z=2$) 
rapidity differ, but all regions preserve useful dependence on dipole moments,
and thus can be useful for the development of observables.
In the following Section we will concentrate on the  $r_{14}$ and $r_{24}$ sensitivity to ${\rm Re}(X)$
and ${\rm Re}(Y)$.


\section{Numerical results on sensitive observable}
\label{sec:decays}

In the previous Section, we have investigated elements of $r_{i,j}$ and $r_{i,4}$ in Eq.~(\ref{eq:rij_definition}).
Now we can use that experience to search for one-dimensional distribution sensitive to anomalous moments which can serve as the prototype for an observable, useful for the tests as well.
Because of neutrinos escaping detections, and because the most sensitive  
to dipole moments turn out to be elements $r_{14}$, $r_{24}$, $r_{13}$ and $r_{23}$,
we use  $\tau^\mp \to \pi^{\mp} \nu_\tau$ decay for its sensitivity to the longitudinal spin,  
and $\tau^\pm \to \rho^{\pm} \nu_\tau \to \pi^{\pm} \pi^{0} \nu_\tau$ decay for its sensitivity to the 
transverse spin. 
In the $pp$ collisions, reconstruction of $\tau$ momenta is difficult and we could only partly
rely on experience of Ref.~\cite{Banerjee:2023qjc}.

Our test observable was defined from the four momenta of the visible decay product system. 
This consists of acoplanarity angle $\phi$ between oriented half-planes built respectively on 
$\vec{p}_{beam1} \times (\vec{p}_{\pi^\pm}+ \vec{p}_ {\pi^0})$ 
and $\vec{p}_{beam1} \times \vec{p}_{\pi^\mp}$.  
Let us provide details of the $\phi$ angle definition.
The {\it beam1} denotes the direction of the first beam~\footnote{Alternatively
we have used also  beam  direction following $z$ component of the visible system 
$\pi^\mp \pi^\pm \pi^0$ momentum.}. 

As sensitivity observable, acoplanarity angle $\phi$ calculated in the rest frame of visible tau lepton
decay products is used. This angle is defined as follows: first we define normal versors $\vec{n}_1$ to the 
plane spanned on the  beam direction and $\pi^{\mp}$ direction of $\tau^{\mp} \to \pi^{\mp} \nu_\tau$ decay, 
and $\vec{n}_2$ for plane spanned on $\pi^{\pm}$ and  $\pi^{0}$ directions for 
$\tau^\pm  \to \rho^{\pm} \nu_\tau$. Acoplanarity angle is then calculated 
as $\arccos$   of $\vec{n}_1 \cdot \vec{n}_2$ scalar product, thus in  $(0, \, \pi)$ range only.
To extend the definition of acoplanarity to one of two oriented half-planes and to $(0, \, 2\pi)$ range,  
the scalar product of $\pi^{\mp}$ from $\tau^\mp$ and $\vec{n}_2$ is used, if it is positive, we  
set acoplanarity angle equal to $2 \pi-\phi$.

Such observable will still have negligible sensitivity to anomalous couplings. We need to  split the 
sample into two parts  according to the sign of the scalar product of 
versor  $\vec{n}_3$  (normal to the plane spanned on the beam versor and $\vec{n}_1$ versor) 
and  $\vec{n}_2$. We denote them as {\tt Region 1} and {\tt Region 2}. 

Fig.~\ref{fig:sens} shows the ratio of acoplanarity angle distribution for three settings of anomalous dipole moments to the SM predictions and for {\tt Region 1} on the left-hand side, and {\tt Region 2} on the right-hand side.

We have chosen for the plots ${\rm Re}(X)=2 \times 10^{-4}$,  ${\rm Re}(Y)=2 \times 10^{-4}$  and imaginary parts of the dipole moments were set to zero, see Fig.~\ref{fig:sens}. Sensitivity to dipole moments through the $\tau$ spin correlations could be demonstrated in semi-realistic condition of observable build from visible decay products of 
$\tau$'s.

The sample used in the analysis has been generated using Pythia 8.2 Monte Carlo generator \cite{Sjostrand:2014zea}, with pp collision at 13.6 TeV c.m. energy and MSTW2008nnlo68cl.LHgrid parametrization of PDFs. 
The sample of $4 \times 10^6$ events of the hard process $q \, \bar q \rightarrow Z \to
\tau^- \tau^+$ was generated with the invariant mass of $\tau$-lepton pair restricted
to the $M_Z \pm 0.5$ GeV range. Tau leptons were decayed with Tauola library \cite{Davidson:2010rw}
without spin correlations included, with one tau decaying $\tau^\mp \to  \pi^{\mp} \nu_\tau$ 
and another tau decaying $ \tau^\pm \to  \rho^{\pm} \nu_\tau \to \pi^{\pm} \pi^{0} \nu_\tau$.
Spin correlations were included with reweighting provided by {\tt TauSpinner} 
package \cite{Przedzinski:2018ett}, to which formulas of \cite{Banerjee:2023qjc} were 
implemented~\footnote{
A comment on the orientation of  reference frames is in place.  
One often interchanges the notations   $\bar{q} q  \to \tau^+\tau^-$ and  $q \bar{q} \to \tau^-\tau^+$.
This seemingly trivial order adjustment is of consequences. In particular for the 
reference frame of the c.m. system. This makes the difference if one  chooses 
the axis $\hat{z}'$  along the $\tau^-$ momentum (as in Section \ref{sec:kinematics}), or takes 
$\bar{q} q \to \tau^+ \tau^-$ and directs $\hat{z}^\prime$ along the $\tau^+$ momentum, as used in 
conventions of {\tt TauSpinner} and also {\tt KKMC} MC. 
Both descriptions are of course equivalent, although to connect different conventions,  rotation around 
the axis $\hat{y}'$ by the angle $\pi$ may be needed as well as interchange of indices $i, j$ 
in the spin-correlation matrix $R_{i, j}^{(q \bar{q})}$. This point was already addressed in 
Ref.~\cite{Banerjee:2023qjc}, see there Section II footnote 1 and Subsection IV.A, in 
context of $e^+e^- \to \tau^+ \tau^-$ production. }. 

Spin weights were calculated for different models: the SM with ${\rm Re}(X)={\rm Re}(Y)=0$, 
and a few versions of anomalous moments included:
(a) with ${\rm Re}(X)={\rm Re}(Y)=2 \times 10^{-4}$, (b) ${\rm Re}(X)=2 \times 10^{-4}, \, 
{\rm Re}(Y)=0$, (c) ${\rm Re}(Y)=2 \times 10^{-4}, \, {\rm Re}(X)=0$.
These dipole couplings are smaller by about a factor of $10^{-3}$ with respect to the SM vector and axial-vector couplings, therefore their linear implementation into 
the differential cross section is reasonable. On the other hand, these dipole couplings are larger than the SM 
value (\ref{eq:mu_SM}) of the anomalous magnetic moment by a factor of $10^2$, so that the chosen values are sufficient to reflect the effect of NP.   

\begin{figure}[!htb]
\vspace{0.5cm}
\begin{center}
  \includegraphics[height=150pt]{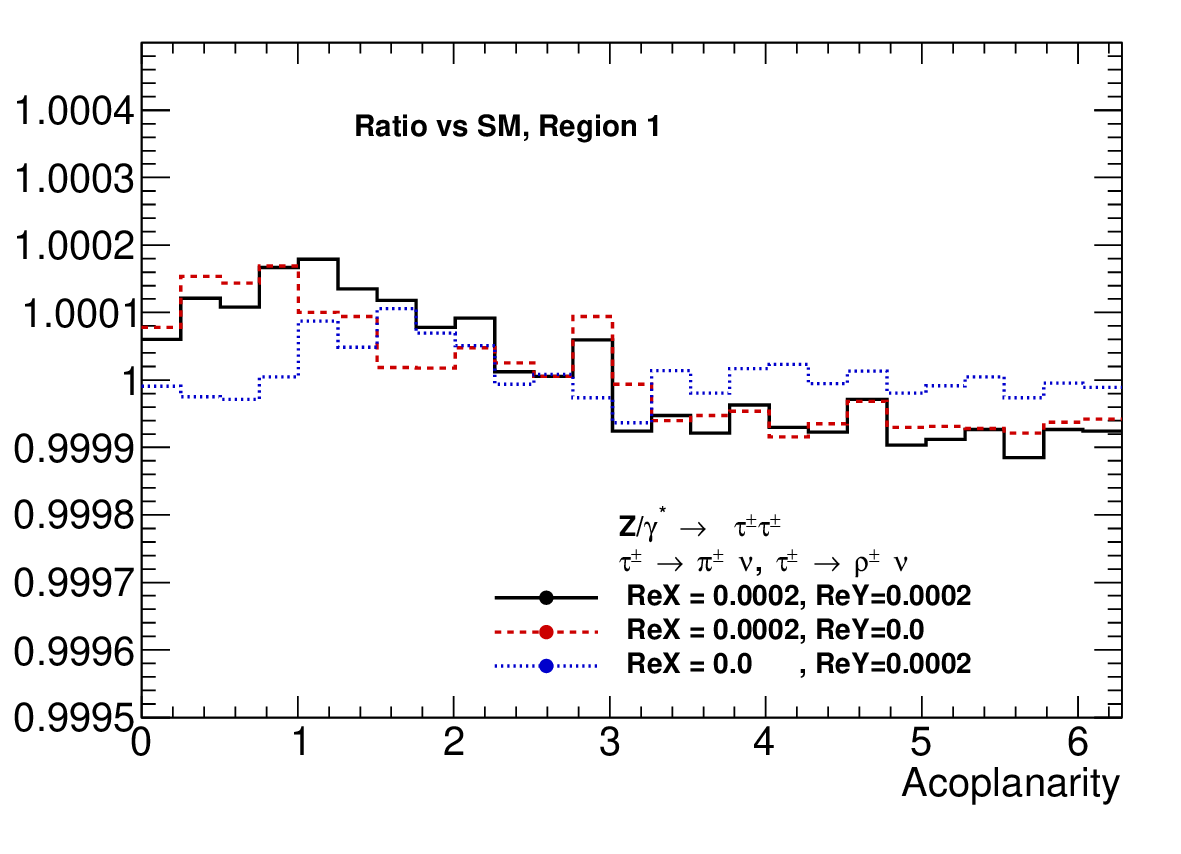} 
    \includegraphics[height=150pt]{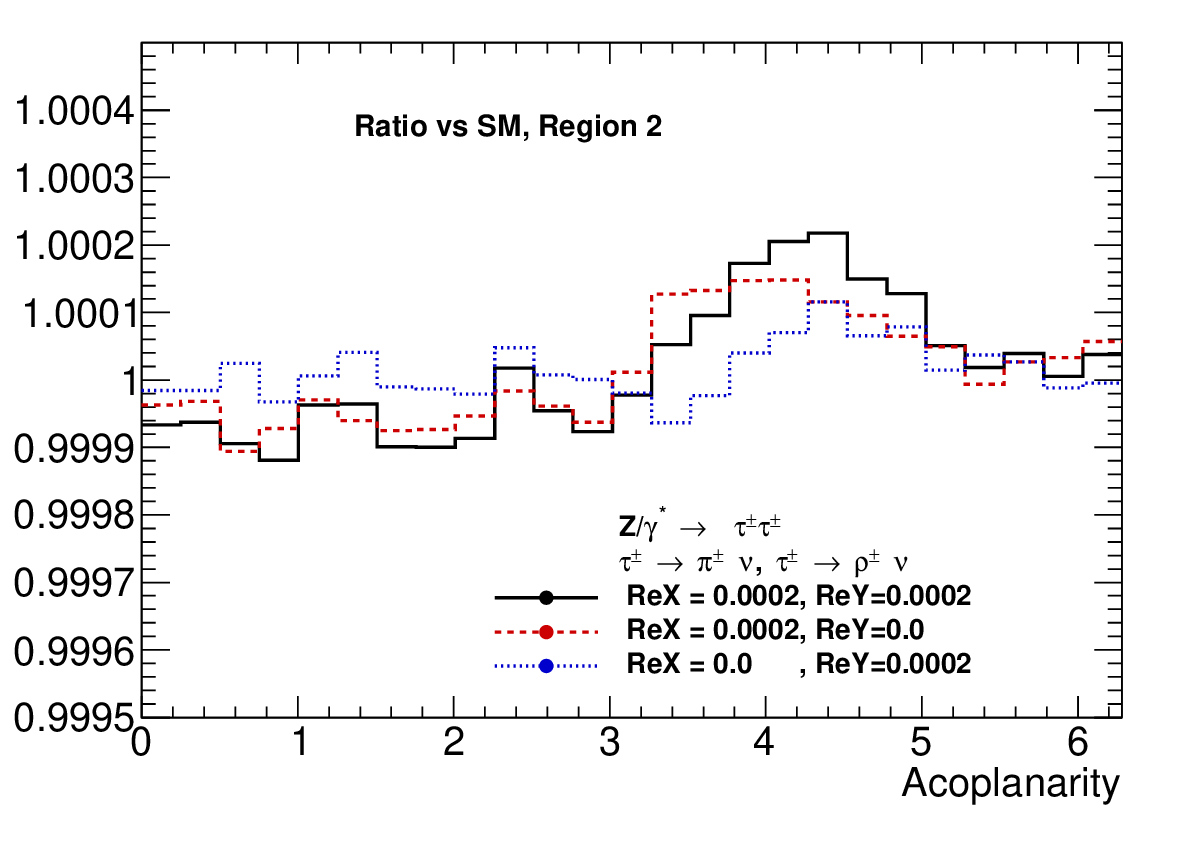}
\end{center}
\caption{Ratio of acoplanarity angle $\phi$ distribution as defined in the text, with: 
(i)  anomalous magnetic, (ii)  electric, and (iii) simultaneously 
 both magnetic and electric dipole moments included, to the distribution in which dipole moments were absent. Selection cut of {\tt Region 1} (left plot) and of {\tt Region 2} (right plot).  }
\label{fig:sens}
\end{figure} 

Fig.~\ref{fig:sens} shows the ratio of acoplanarity angle distribution with/without anomalous dipole moments. Distributions in {\tt Region 1} and {\tt Region 2} are shown separately. One can observe the effect on the shape 
of distribution on the level of $10^{-4}$ in the {\tt Region 1} and {\tt Region 2} for the acoplanarity angles 
$(0, \, \pi)$ and $(\pi, \, 2 \pi)$, respectively.      

As the dipole moments are introduced at the linear level  through the  interference with the SM amplitudes,
their effect shown in Fig.~\ref{fig:sens} can be obtained for other values of $X(M_Z^2)$ and $Y(M_Z^2)$
by a simple rescaling. Note also, that because of weighted event samples, the statistical error affects
the NP effect alone and  not from the difference of  the independently obtained results for
distinct anomalous coupling values. Therefore, the
obtained numerical results are meaningful even though a sample of only $N=4 \times 10^6$ events was used.
That is why, statistical error on NP effect is substantially smaller than   $1/\sqrt{N} = 0.5 \times 10^{-3}$
expected for the compared distribution features.

 In future we can use $\pi^\mp$ energy, or equivalent variable, to explore anomalous moment dependence
due to elements $r_{13}$ and $r_{23}$ which are sensitive to imaginary parts of  the weak dipole moments 
$X(M_Z^2)$ and $Y(M_Z^2)$.

\section{Summary}
\label{sec:summary}

We have investigated the possibility of using the spin effects of the Drell-Yan $\tau$-pair production and decay in the 
LHC conditions. For that purpose we constructed observables sensitive to the presence/absence of anomalous dipole moments of the $\tau$-lepton interaction. To simplify the task
we have concentrated on couplings to the $Z$ boson. Interaction with virtual photons is potentially more 
involved as no constraint of the $Z$ peak can be used. The couplings to virtual photons can be measured 
in the Belle experiments, for relatively low $\tau$-pair virtuality though. 
For higher virtuality more complex evaluation 
of experimental cuts than presented here, would be needed. We leave it to the forthcoming work, 
preferably by the experiments themselves as discussion of
experimental cuts will be of much greater importance.

We expect that our observable, demonstrating the sensitivity of spin-dependent observables to anomalous dipole moments, is primarily for the program tests.
In final applications, Machine Learning (ML) techniques will be far more suitable, but 
even in those cases, idealized optimal-like variables may be helpful to prepare the primary set of variables 
used as input for ML algorithm training. Such variables may help study experimental ambiguities too.
The demonstrated extension of {\tt TauSpinner} algorithm may be helpful in such projects.

\section*{Acknowledgment}

Numerical calculations were partially performed at the PLGrid Infrastructure of the Academic Computer
Centre CYFRONET AGH in Krakow, Poland.
This research was funded in part by Narodowe Centrum Nauki, Poland, grant No.~2023/50/A/ST2/00224.

\bibliographystyle{unsrt}
\bibliography{p-p_tau-tau}

\end{document}